\documentclass[preprint2]{aastex}
\usepackage{epsfig}

\newcommand{\he}{Hen~3-1475\ }
\newcommand{\hhe}{Hen~3-1475}

\shorttitle{Polarization and IR photometry of Hen~3-1475}
\shortauthors{Rodrigues et al.}

\begin{document}


\title{Optical polarization and near IR photometry of the proto-planetary nebula
Hen~3-1475\footnote{Based on observations made at Laborat\'orio Nacional de
Astrof\'\i sica/MCT, Brazil}}


\author{Cl\'audia V. Rodrigues\altaffilmark{2}, Francisco J.
Jablonski\altaffilmark{2}, Jane Gregorio-Hetem\altaffilmark{3},
Gabriel R. Hickel\altaffilmark{2,4} and Mar\'\i lia J.
Sartori\altaffilmark{5,3}}

\altaffiltext{2}{Instituto Nacional de Pesquisas
Espaciais/MCT --- Av. dos Astronautas, 1758 --- 12227-010 - S\~ao Jos\'e dos
Campos - SP --- Brazil}
\altaffiltext{3}{Instituto de Astronomia, Geof\'\i sica e Ci\^encias
Atmosf\'ericas/USP --- Rua do Mat\~ao, 1226 --- 05508-900 - S\~ao Paulo -
SP --- Brazil}
\altaffiltext{4}{Universidade do Vale do Para\'\i ba - IP\&D ---
 Av. Shishima Hifumi, 2911 ---
 12244-000 - S\~ao Jos\'e dos Campos - SP --- Brazil}
\altaffiltext{5}{Laborat\'orio Nacional de Astrof\'\i sica/MCT --- Rua Estados Unidos, 154
--- 37504-364 - Itajub\'a - MG --- Brazil}

\email{claudia@das.inpe.br}

%


\begin{abstract}

We present BVRI CCD aperture polarization and near-infrared photometry of the
proto-planetary nebula Hen 3-1475. Its intrinsic polarization is high and shows
a strong spectral dependence. The position angles in all bands are
perpendicular to the axis of the observed bipolar structure. A Monte Carlo code
is used to model the intrinsic polarization of \hhe. Using disk dimensions and
other constraints suggested by previous works, we are able to reproduce the
observations with an optically thick disk composed by grains
with a power-law size distribution ranging from 0.06 to 0.22 $\mu$m.
We also reliably estimate the
foreground polarization from hundreds of stars contained in the CCD images. It
is parallel to the intrinsic polarization of Hen 3-1475. Possible
implications of this result are discussed. From IR observations, we estimate a
interstellar reddening, A$_V$, of about 3.2. 

\end{abstract}




\section{Introduction}

\he (IRAS 17423-1755) is identified as a proto-planetary nebula (PPN),
i.e., an object between the end of the asymptotic giant branch
(AGB) and the planetary nebula (PN) phases. The classification
and physical properties of Hen~3-1475 are supported by a broad base of
observational data: radio emission  (Knapp et al. 1995); OH masers (te Lintel
Hekkert 1991; Zijlstra et al. 2001); optical imaging and spectroscopy (Riera et
al. 1995; Bobrowski et al. 1995; Borkowski, Blondin, \& Harrington 1997;
Borkowski \&
Harrington 2001). Probably the strongest evidence in favor of its evolutionary
stage is the high N abundance, typical of evolved objects, specifically of
Type I PN (Riera et al. 1995). The radio continuum indicates the presence of a
compact ionized region  surrounding the central B star with an estimated
dynamical age of only 15 years (Knapp et al. 1995. Recently, Borkowski \& Harrington
(2001) have
obtained images and detailed spectroscopic observations and, among other
results, they have  derived a distance of 8~kpc, which implies a very high
luminosity (25\,000 L$_\odot$) for a AGB star at the Galactic Bulge.
Analyzing H$\alpha$ spectra, S\'anchez Contreras \& Sahai (2001) could
identify two winds, one of which is probably associated with the post-AGB wind
before its interaction with the previous AGB wind.

High-resolution optical images of Hen~3-1475 show a very complex bipolar
morphology (Borkowski et al. 1997; Bobrowski et al. 1995; Riera et al. 1995).
The central region is dense, probably a dust disk, which may be responsible
for the OH maser emission as well as for the CaII and FeII lines observed in its
near-infrared spectrum. This structure has an angular size of about 2\arcsec.
The high resolution image obtained by Borkowski et al. (1997) shows also dark
patches probably associated with denser clumps of material in the disk. 
A complex structure of highly collimated optical
jets with pairs of point-symmetrical shock-excited knots produces a characteristic
S-shaped sequence, which is about 15\arcsec\ in size oriented along PA 135$^o$.

In the rapid evolutionary phase of PPN, dramatic changes in geometrical and
physical
properties of its central star and circumstellar material take place.
An example is the development of non-symmetric geometries (eg., bipolar
nebulae) in post-AGB objects. The interaction between an equatorial enhanced
AGB wind and that from an early PN phase was suggested as a mechanism to
produce the variety of geometries seen in PPNe and PNe (Kahn \& West
1985). The status of this and other models is reviewed by Frank (2000).  This
enhanced equatorial wind - which we will call disk - can also play a role in
collimating the jets seen in many objects. Recently, Frank, Lery, \& Blackman
(2002) have shown that a magnetic-centrifugal launching model can explain winds with
velocities as high as those observed in \hhe.  Therefore, a good representation
of the disk properties is crucial to understand the changes
occurring during the PPN phase.

Polarimetry is an useful tool in the study of PNe and PPNe because it is
sensitive to the circumstellar geometry and composition. The polarization
measured in PPNe is usually high due to the scattering in
asymmetric circumstellar environments. Johnson \& Jones (1991) have shown that,
from red giants to PNe, the evolutionary stage of PPN is the one in which the
polarization tends to be highest and that the transition from a spherically
symmetric envelope (typical of red giants) to a non-spherical envelope seems to
occur in early stages of the AGB phase. The wavelength dependence of the
polarization in PPNe has various shapes and some objects present
polarization-angle rotation (Trammell, Dinerstein, \& Goodrich 1994). Many
extended objects were observed using imaging polarimetry (e.g., Scarrot \&
Scarrot 1995; Gledhill et al. 2001). Far from the central source, the light is
highly polarized and exhibits the centro-symmetric pattern typical of reflection
nebulae. However, the polarization near the central object tends to be parallel
to the optically thick disk originated in the AGB phase. This pattern is also
observed in young stellar objects with bipolar geometries (see references in
Bastien \& Menard 1988 - Table 1).

In spite of the importance of grain size in the study of the temperature
profile and, consequently, of the infrared (IR) luminosity (Stasi\'nska \&
Szczerba 1999, 2001), few works have focused on the determination of the grain
size distribution in PPNe and PNe. An example of model that considers the
grain sizes as free parameters is found in Meixner et al. (2002). That work,
however, modeled the
observed spectral energy distribution (SED) alone. Recent results (Carciofi,
Bjorkman, \& Magalh\~aes 2002; Li \& Lunine 2002) show that the SED alone does
not constrain the grain sizes. The spectral dependence of the polarization
helps to restrict the models to the dust size distribution (Carciofi, Magalh\~aes, \&
Bjorkman 2002). Heckert \& Smith (1988) modeled the polarization
of OH 0739-14 considering single scattering in an envelope composed by grains of
a single size. Johnson \& Jones (1991) have modeled the observed polarization
of some PPNe, but they used a fixed MRN size distribution ranging from 0.005 to
0.10 $\mu$m. To our knowledge, there is no previous attempt to describe the
polarization data of a PPN with a free distribution of grain sizes and
multiple scattering.

In this work, we present data on broad band CCD aperture polarimetry as well
as near-infrared (NIR) photometry of Hen~3-1475 (Section \ref{sec_obs}).
In Section \ref{sec_ir} the NIR photometry is analyzed in order to provide
information on the interstellar (IS) and intrinsic reddening. The polarimetric
observations are discussed and modeled in Section \ref{sec_polar} providing some
information about the Hen~3-1475 dusty envelope. In the last section, the main
conclusions are listed.

\section{Observations}
\label{sec_obs}

\subsection{Optical Polarization}

The polarimetric observations have been done with the 0.60-m Boller \& Chivens
telescope at Observat\'orio do Pico dos Dias, Brazil, operated by
the Laborat\'orio Nacional de Astrof\'\i sica (LNA), Brazil, using a
CCD camera modified by the polarimetric module described in Magalh\~aes et al.
(1996). The CCD array used was a SITe back-illuminated, $1024 \times 1024$
pixels. The above telescope and instrumentation gives a field-of-view of
10\arcmin.5 $\times$ 10\arcmin.5. The polarization in the V band was 
measured on 1999 April 11 and July 29; the observations in the (RI)$_C$ bands
were also obtained on 1999 July 29 and in the B band, on 1999 July 30. In each
night, we observed polarimetric standards stars (Serkowski, Mathewson, \&
Ford 1975; Bastien et al. 1988; Turnshek et al. 1990) in order to calibrate
the system and estimate the instrumental
polarization. The measured values of the unpolarized standard stars were
consistent with zero within the errors: consequently no instrumental correction
was applied.
Measurements using a Glan filter were also performed to estimate the efficiency
of the instrument. They indicate that no instrumental correction is needed.

The images have been reduced following the standard steps of differential
photometry using the IRAF\footnote{IRAF is distributed by
National Optical Astronomy Observatories, which is operated by the Association
of Universities for Research in Astronomy, Inc., under contract with the
National Science Foundation.} facility. Counts were used to
calculate the polarization using the method described in Magalh\~aes,
Benedetti, \& Roland (1984). The measured polarization of \he is shown
in Table \ref{tab-pol}. The aperture radius that minimizes photometry errors is
4 pixels (corresponding to 2.44$\arcsec$ in the sky). It was kept 
constant for all bands. In the Discussion and Conclusion sections we will have in
mind that the polarization measurements refer to an extended region in the
vicinity of the central object in \hhe.
The multicolor polarimetry was obtained in nights of poor photometric conditions
with high sky counts. Although this does not have important consequences on
polarization results, which are based on a differential technique, this
precludes us from detecting the nebula surrounding the central object.

\subsection{IR Photometry}

Hen 3-1475 was observed with the CamIV NIR imager in the 1.6-m telescope of
LNA on 1999 July 29. The detector is a HAWAII array, with 0.24\arcsec/pixel
spatial resolution, covering 4\arcmin $\times$ 4\arcmin. J (1.25$\mu$m) and H
(1.65$\mu$m) band images were obtained. This region is already available in the
Two Micron All Sky Survey (2MASS) survey, which we used to calibrate our
photometry. Our combined images have an effective exposure time of 210 s in J
and 40 s in H and provide S/N $\approx$ 10 for H=17. This is deeper than the
2MASS photometry at the same S/N,  also with a better spatial resolution,
allowing us to discuss the gross features of the reddening in the line-of-sight
to Hen 3-1475. The J and H magnitudes for the target were obtained from short
exposure time images and are $J = 9.75 \pm 0.01$ and $H = 8.44 \pm 0.01$.

\section{Inter- and Circumstellar Reddening of Hen 3-1475}
\label{sec_ir}

We have obtained the magnitudes of 213 objects around Hen 3-1475. An examination
of the
color-magnitude diagram (CMD) H $\times$ (J-H) allows us to put limits on the
total reddening to this line of sight, since the distribution of unreddened
stars is relatively localized. Hen 3-1475 has $l = 9.8\arcdeg, b = +5.3\arcdeg$,
corresponding to a line-of-sight that intercepts a relatively dense part of the
bulge of the Galaxy. The objects we observe with H $<$ 14 and (J-H) $\approx
+0.8$ probably are red giants in the bulge. If they were not subject to
reddening, the distribution of their color indices should have a maximum around
J-H $\approx +0.5$. This can be obtained by a simple implementation of the model
of Wainscoat et al. (1992). Using the  Cardelli, Clayton, \& Mathis (1989)
parametrization of the IS extinction curve, we obtain E(J-H) = 0.09
A(V), for R = 3.1. So we conclude that the reddening to the red
giants of the bulge is A(V) $\approx 3.2$ mag in this line-of-sight. If R = 5,
A(V) would change to 2.8 mag. These values are a bit larger than the previous
estimate of Riera et al. (1995), A(V) $<$ 2.0 $\pm$ 0.4 mag.

The total reddening to \he presents a much more complex problem, since
the object is shrouded by circumstellar material.
We assume an spectral type of B3 (Knapp et al. 1995). The object
position in the CMD is not consistent with that of a dwarf, rather, a
supergiant star is needed. Considering the color of a B3I (Tokunaga 2000), we can
estimate a circumstellar E(J-H) of 0.84 after subtracting the interstellar
value. Converting the derived E(J-H) to A(V) is not straightforward. 
A first guess could be made using a standard interstellar extinction law: this gives
an intrinsic A(V) = 9.1 (7.9), if R = 3.1 (5.0). However, the
spectral energy distribution of \he may be quite different from that of a blue star
attenuated by interstellar 
extinction law. At optical wavelengths the presence of 
scattered light - a blue component - may be important and at NIR wavelengths hot
grains could have a non-negligible contribution to the spectrum.
A model taking in account all these components will be
explored in a forthcoming paper.

\section{Discussion on polarimetric results}
\label{sec_polar}

In this section, we present the estimates to the foreground and intrinsic
polarization of \hhe. We also discuss a probable alignment between the
intrinsic polarization of this object and the interstellar magnetic field.
Finally, numerical modeling of \he polarization is done assuming it is
produced in a circumstellar disk.

\subsection{Intrinsic polarization of Hen~3-1475}

The observed polarization in astrophysical objects is usually the sum of
two components: (i) the intrinsic one, produced in the object itself; (ii)
the foreground component, which is produced in the interstellar medium (ISM)
between the object and the Earth. The high polarization of \he suggests that at
least part of it may have an intrinsic origin because it is very unusual such a
high value ($\approx$ 8\% in V band) to be produced in the ISM. To obtain the
intrinsic component, it is necessary to estimate the foreground polarization and
subtract it from the observed one.

\begin{figure}
\plotone{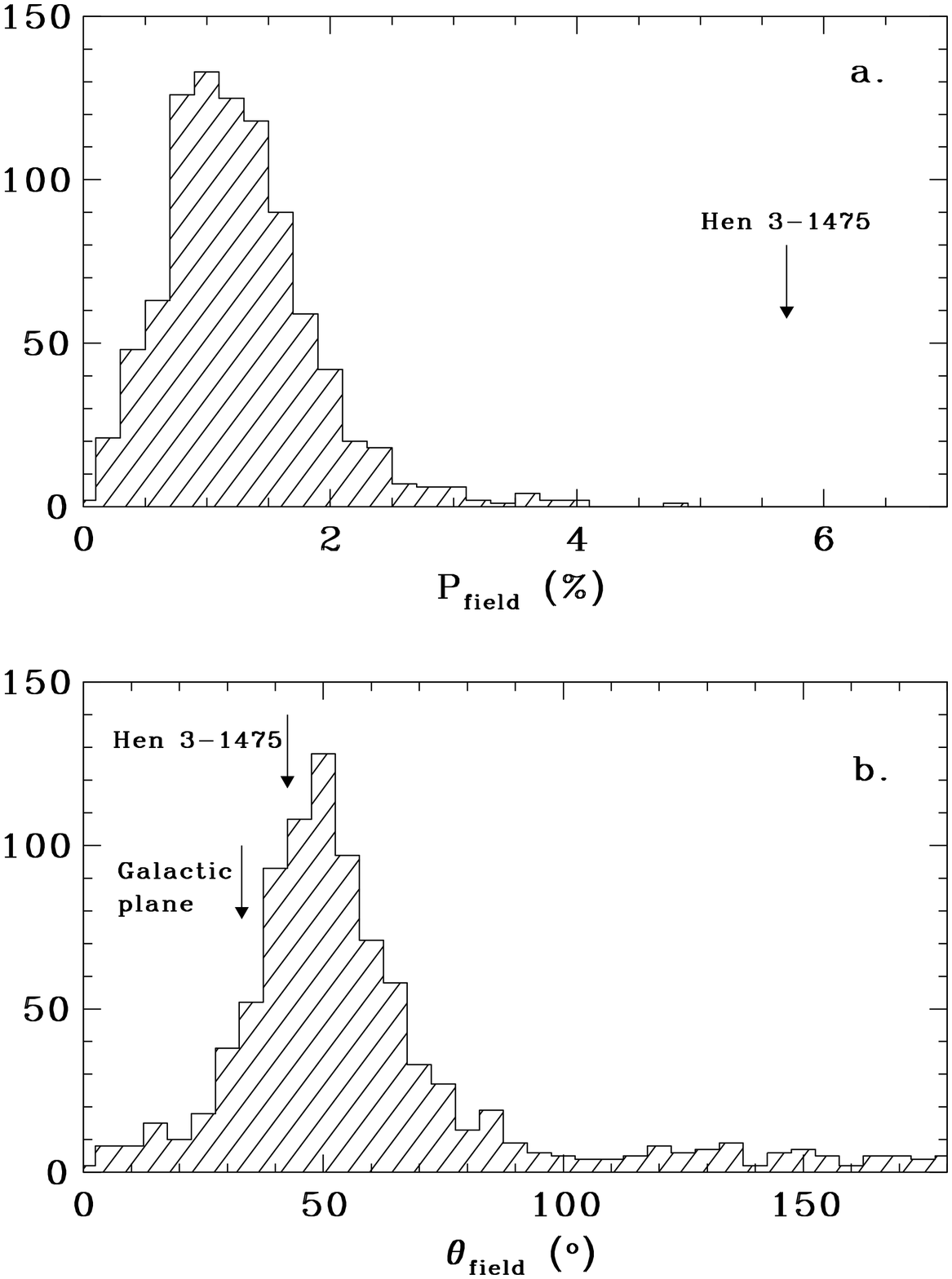}
\caption{Histograms of the field stars polarization in I$_{c}$ band: a.
Polarization degree; b. Position angle.}
\label{fig-foreg}
\end{figure}

The foreground polarization in each band was estimated by the weighted
average polarization of all stars in the field-of-view. The results are
presented in Table \ref{tab-pol}. As the line-of-sight to \he is
near the Galactic center ($ l = 9.8^o ; b = +5.3^o$), the images contain a
large number of stars (see last column of Table \ref{tab-pol}). To illustrate
the distribution of the polarization of the field stars we present
histograms for their moduli and position angles in I$_C$ band (Fig.
\ref{fig-foreg}.a and Fig. \ref{fig-foreg}.b): notice that the maximum of each
histogram coincides with the estimated values of the foreground polarization
(Table \ref{tab-pol}). The histograms show single-mode distributions consistent
with a sample of stars coming from a singular population with regard to
(interstellar) polarization properties. This may be interpreted as an evidence
that the foreground dust shares the same properties all over this field.
The histogram in Fig. \ref{fig-foreg}.a also shows how large the polarization of
Hen~3-1475 (arrow) is in comparison with the field stars. This reinforces the
suspicion that this object has a large intrinsic component.

Another evidence to the intrinsic character of the \he polarization comes from
its wavelength dependence. We have fitted both the observed Hen~3-1475 polarization
and that of the field stars with the Serkowski law (1973):

\begin{equation}
P(\lambda) = P_{max} \exp{\left\{ -K
\ln^2\frac{\lambda_{max}}{\lambda}\right\}} ,
\end{equation}

\noindent where: $P_{max}$ is the maximum value of the polarization;
$\lambda_{max}$ is the wavelength where $P_{max}$ occurs; and $K$ is a
parameter related to the curve width and dependent on $\lambda_{max}$ 
(Serkowski, Mathewson \& Ford 1975; Codina-Landaberry \& Magalh\~aes
1976; Wilking, Lebofsky, \& Rieke 1982). In the fits, we have used 
$K = 1.66 \lambda_{max} + 0.01$ (Whittet et al. 1992).

\begin{figure}
\epsfig{file=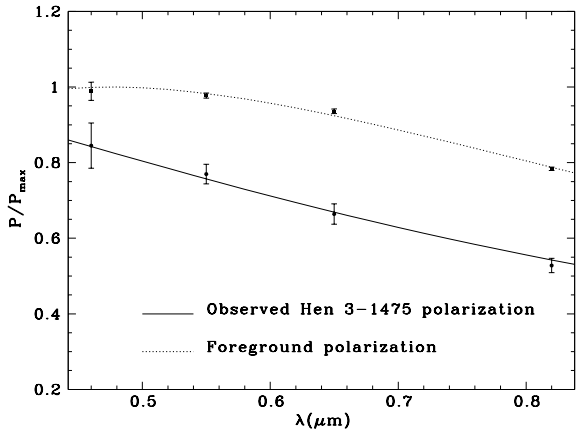,width=6cm,angle=-90}
\caption{Polarization data and Serkowski fits to the polarization of Hen
3-1475 and field stars.}
\label{fig-serk}
\end{figure}

The results are shown in Table \ref{tab-serk} and Figure \ref{fig-serk}:
the polarization was normalized to $P_{max}$ to
allow an easier comparison between the curves. In this graph it is clear that
the wavelength dependence of the polarization of \he and
the field stars are different. Also, $\lambda_{max}$ occurs at a short
wavelength for \he when
compared to the behavior of the field stars. Such a short value would be quite unusual
if its origin was interstellar.

Based on the discussion above, we are confident that \he has an intrinsic
polarization produced in its circumstellar region. It was estimated subtracting
the average polarization of the field stars from 
that observed for \hhe. These values are presented in Table \ref{tab-pol}. If we estimate
the foreground polarization
from a Serkowski law fit with a constant position angle, the changes in the intrinsic
polarization are negligible. Hereafter, the
term ``\he polarization" refers to its intrinsic value.

\subsection{Geometry of Hen~3-1475 and its relation with the ISM}

The Hen~3-1475 jets are aligned at approximately PA 135$^o$
(Riera et al. 1995; Bobrowski et al. 1995). The PA of the
polarization, $42.1^o$ (in V band), is consistent with the
observed bipolar geometry: it is perpendicular to the jets and parallel to the
dust torus. This behavior was already observed in young stellar  objects with
bipolar geometry (Hodapp 1984; Sato et al. 1985). The mechanism causing this
polarization will be discussed in the Section \ref{sec_model}: in this section
we would like to discuss a possible correlation of the geometry of \he
with the orientation of the magnetic field of the surrounding ISM.

Curiously, the intrinsic polarization of Hen~3-1475 is nearly parallel
to the polarization of the field stars ($48.3^o$ in V band). The latter effect is
usually attributed to the transport of starlight through a dichroic medium.
This dichroism is caused by non-spherical grains aligned by the interstellar
magnetic field. Considering the standard model for grain alignment in the
ISM, the polarization angle is parallel to the component of the magnetic field
in the plane of the sky (Davis \& Greenstein 1951; but see also Lazarian
2000). The foreground polarization direction in the field of Hen~3-1475 is
almost parallel to the Galactic plane, which is at PA 33$^o$. This result is
consistent with those from maps of optical polarization that indicate that the
direction of the magnetic field in the Galaxy follows the
Galactic plane (Mathewson \& Ford 1970; Axon \& Ellis 1976).

The alignment between \he geometry and field stars
polarization can be interpreted as: (i) a chance effect; (ii) the foreground
polarization is caused by a cloud physically near \he and in this case the
IS magnetic field aligning the grains is also present in the \he environment;
(iii) the cloud producing the polarization of field stars is far from \he but its
magnetic field is aligned with the Galactic plane, so the alignment of \he may
be regarded as relative to the Galactic plane. 

Considering objects of similar evolutionary status, it is not clear if there
is a correlation between the nebula axis and the Galactic plane. Recent results
indicate that the symmetry axis of PNe are randomly distributed (Corradi,
Aznar, \& Mampaso 1998). However, this work excluded PNe with small scale
structures as the point symmetrical knots seen in \hhe, for instance. On the
other hand, previous works had suggested some correlation between PNe axis and
the Galactic plane (Melnick \& Harwit 1975; Phillips 1997). Nebulae ejected by
massive stars and bilateral supernova remnants seem to have their axis of
symmetry along the Galactic plane (Hutsemekers 1999; Gaensler 1998). Possible
mechanisms that could cause such a correlation are discussed in Melnick \&
Harwit (1975) and Phillips (1997).

Another possibility is that the axis of symmetry of \he is determined by
the IS magnetic field.
The dust disk of Hen~3-1475 is probably a remnant of the AGB wind. It is
quite unlikely that the IS magnetic field could have an important role in
causing the asymmetry of the dense AGB wind\footnote{Dgani (1998) has done some
calculations showing that filamentary structures seen in old (=low density) PN
can be explained on basis of their interaction with the magnetic field of the
ISM.}. Alternatively, the connection between an AGB stellar wind asymmetry and
ISM might be a fossil from the star formation phase. Massive stars can indeed
easily retain the main-sequence angular momentum (Garcia-Segura et al. 2001). 
We briefly present some results in the literature concerning the alignment of
young stellar objects geometry with the IS magnetic field. This is still
a mostly open subject in which studies using different techniques provide
discordant results. Dyck \& Lonsdale (1979) are among the first authors
to notice a possible alignment between the direction of the IR
polarization of protostellar sources and the IS magnetic field. Heckert \& Zeilik (1981)
show that IR polarization of YSOs tends
to be parallel or perpendicular to the IS magnetic field.
The large scale molecular outflow tends to be aligned with the IS magnetic field
(Snell, Loren, \& Plambeck 1980; Hodapp 1984; Cohen, Rowland \& Blair, 1984;
Vrba et al. 1986). Recent sub- and millimetric polarimetric studies of prestellar
objects are not yet conclusive whether there is (or not) a correlation between
the core elongation/outflow and the magnetic field of the containing filament
(Ward-Thompson et al. 2000; Matthews \& Wilson 2000; Glen, Walker, \& Young
1999). 

\subsection{Modeling the circumstellar environment of Hen~3-1475}
\label{sec_model}

To model the spectral dependence of the polarization in \hhe, we have used
a three dimensional Monte Carlo code for radiative transfer in stellar
envelopes (Rodrigues 1997; Rodrigues \& Magalh\~aes 2000).
It includes scattering and absorption by dust particles, but their
emission is neglected: therefore all the photons are originated in the central
source. As we are modeling optical emission this can be considered as a
reasonable approximation. The grains are considered spherical and homogeneous.
We have used three optical constants in the models: astronomical silicate
(Draine \& Lee 1984 - DL94); O-rich silicate from Ossenkopf, Henning, \& Mathis
(1992 - O92); and amorphous carbon from Zubko et al. (1996 - Z96). The phase
function considered is not the Henyey-Greenstein approximation, but the
analytical function obtained from the Mie theory. The density, $\rho(r)$, was
assumed to depend on the distance as $r^{-2}$.

The diameter of the \he disk torus is about 2$\arcsec$ (Borkoswki et al. 1997).
It
was suggested that the jets are collimated near the first pair of knots, which is
located around 3$\arcsec$ from the central object (Borkowski et al. 1997).
Therefore the aperture radius used in the polarization measurements
includes the disk, but not the jets. So, the intrinsic polarization, as obtained
in this work, should originate mainly by processing of the central source light
in the optically thick disk torus. This configuration can indeed produce a
polarization parallel to the disk as observed in objects with bipolar geometry
(Bastien \& Menard 1988; Hodapp 1984; Sato et al. 1985). This can be
explained by multiple scattering in an optically thick disk (Bastien \& Menard
1988, 1990; Whitney \& Hartmann 1993; Fischer, Henning, \& Yorke 1994).
We have adopted a cylindrical geometry to represent the \he disk.

The models have a relatively large number of input parameters
(Table \ref{tab-input}). In order to minimize the number of degrees of freedom,
we have fixed a given parameter if it has a reasonable good published estimate - see
in Table \ref{tab-input} whether a parameter was considered fixed, its
value and the reference to the estimate. If \he is at 8 kpc (Borkowski \&
Harrington 2001), 1$\arcsec$ corresponds to a physical size of 0.04
pc ($1.7 \times\ 10^6$ $R_{\odot}$). This value was used for the external radius
of the disk. Its inner radius was estimated to be about 0.01 pc = 4 $\times\ 10^5\
R_{\odot}$ according to the IRAS fluxes (Bobrowski et al. 1995). This is
consistent with simulations of formation of PN envelopes (Mellema \& Frank
1995). To estimate the stellar radius we have considered a spectral type B3
(Knapp et al. 1995) and a luminosity of 25\,000 L$_\odot$ (Borkowski \&
Harrington 2001), which results in a value of 20 R$_\odot$. In a
cylindrical geometry, there is one more parameter: the disk height, $h$, which
we varied. We may notice that the results are invariant if
all lengths are multiplied by a given factor and if the optical depth, $\tau$, is
kept constant. We can also fix the disk inclination, adopting the inclination
angle of the bipolar outflow obtained from Borkowski \& Harrington (2001), which
is 40$\arcdeg$. 

Before analyzing the quantitative results for \hhe, we discuss
how each parameter affects the polarization curve. In the optically
thick regime, the optical depth controls basically the polarization degree and
slightly the shape of the spectral polarization curve.  
The polarization curve obtained by a model with a distribution of grain sizes
has a maximum at a wavelength related to the average grain size.
The width of that curve is related to the
width of the grain size distribution. In Hen 3-1475 we do not see the
polarization maximum, but it may occur at smaller wavelengths than those
covered by our observations. The cylinder
height, $h$, plays a role that is not so obvious. For the present case, when we
consider a fixed line of sight (= inclination), the height defines whether the
central source light is blocked or not by the envelope: the more direct
starlight, the smaller the polarization degree. Also, a thicker cylinder
contains more matter, which also contributes to increase the polarization.
Hence, the cylinder height strongly affects the polarization degree. Concerning
the \he polarization model, a small height allows the central object to be seen
directly, producing a polarization that is not high enough. We conclude that the
minimum cylinder height that is needed to fit the polarization is $h_{min} \approx 7.
\times 10^5 R_\odot$.

We have initially tried to fit the wavelength dependence of the
\he polarization using a single grain size but the width of the obtained curves were
too narrow, so a size distribution had to be introduced.
We have used the recipe of
Mathis, Rumpl, \& Nordsieck (1977 - MRN), the standard model for the
interstellar medium grains. It is represented by a power-law, $a^q$, with
adjustable limits, $a_{min}$ and $a_{max}$.

In order to fit the \he polarization data, we have proceeded in the following way:
we fixed the parameters related to the envelope geometry and
the index of the grain-size power-law distribution, q. Then, we let the
the grain size distribution limits and the optical depth vary in order to find
the best model in a weighted chi-square sense. The best fits are shown in
Table \ref{tab-model}.

\begin{figure}
\epsfig{file=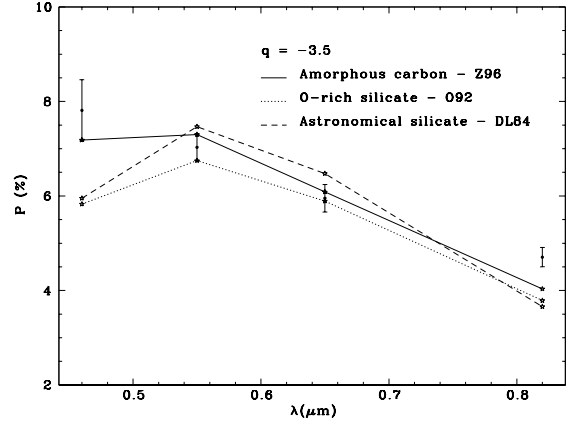,width=6cm,angle=-90}
\caption{Models to the intrinsic polarization of \he using different grain
composition.}
\label{fig_mod_chem}
\end{figure}

\begin{figure}
\epsfig{file=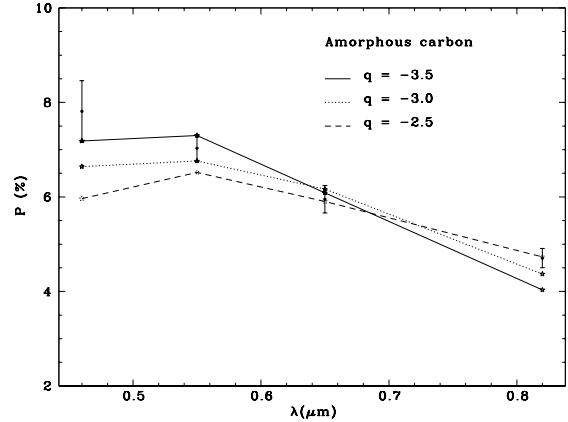,width=6cm,angle=-90}
\caption{Models to the intrinsic polarization of \he varying the index of
the power-law size distribution.}
\label{fig_mod_q}
\end{figure}

\begin{figure}
\epsfig{file=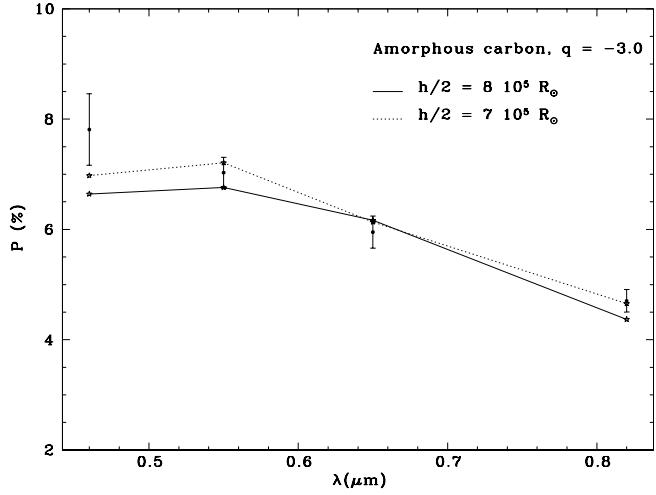,width=7cm,angle=-90}
\caption{Models to the intrinsic polarization of \he for different cylinder
heights.}
\label{fig_mod_h}
\end{figure}

Let's initially suppose a grain size distribution that has an
exponent, q, of -3.5. In spite of having an oxygen-rich envelope (Knapp et
al. 2000), the optical properties of amorphous carbon provide a better fit
to \he polarization data than the silicate ones. This is illustrated in Table
\ref{tab-model} (see models 1, 5 and 7) and also in Fig. \ref{fig_mod_chem}.
This graph also shows that all models tend to produce polarization curves 
narrower than the observed. An improvement could
be in principle obtained using less steep power-law grain size distributions.
Varying the $q$ index for an amorphous carbon model (see models 1, 2 and 3 and also Fig.
\ref{fig_mod_q}), we could improve the chi-square (best fit for $q$ = -3.0), but
the shape of the polarization spectral curve was not well described yet. An
improvement could
also be achieved by decreasing the cylinder height (see model 4 and Fig.
\ref{fig_mod_h}). The grain sizes of the models in Table \ref{tab-model}
do not differ significantly, in spite of
being obtained using models with different values of $q$ and $h$:
so the dust sizes are relatively well-constrained. Considering the best fit (model 4), the
polarization of \he can be fitted by amorphous carbon grains with size in the
range 0.063 - 0.22 $\mu$m
following a -3.0 power-law. We notice, however, that silicate is not
discarded as a possible dust composition. A more refined model should be used
to better constrain the chemical composition (see below). Table
\ref{tab-model} shows that the grain sizes are moderately affected by
the optical constants.

\begin{figure}
\epsfig{file=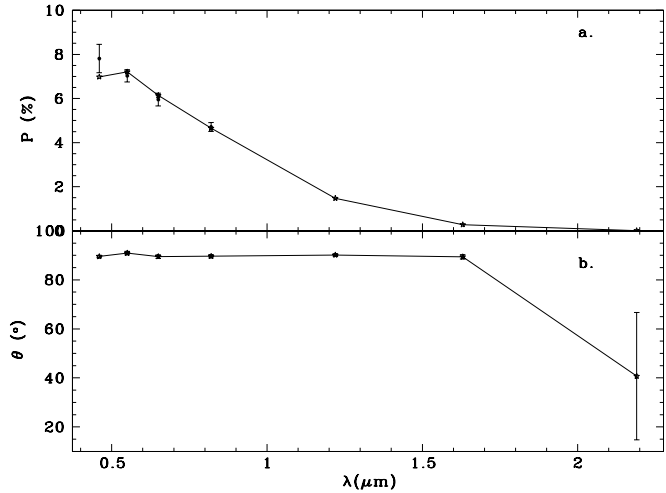,width=7cm,angle=-90}
\caption{Polarization (a) and position angle (b) in the BVRIJHK
bands for the best fit. The error bars in panel b are estimates
to statistical error of the model.}
\label{fig-jhk}
\end{figure}

As mentioned above the measured polarization of \he is parallel to the
dust torus (see Table \ref{tab-pol}).  Fig. \ref{fig-jhk} shows the degree and
the position angle of the polarization obtained with the model 4. In the
position angle panel a value of 90\arcdeg\ corresponds to a polarization
parallel to the disk: this means that the BVRI observations are reproduced.
However, the optical depth decreases to the red and in some point the envelope
becomes optically thin, so the polarization could become perpendicular to the
disk. This would produce a behavior similar to the Type 1c objects from Trammell
et al. (1994): a rotation of 90\arcdeg\ and a decreasing polarization from small
(optically thick) to large wavelengths (optically thin). To check that
possibility, we have calculated the JHK polarization produced by model 4, which
is also shown in Fig. \ref{fig-jhk}\footnote{The position angle in the K band
obtained using a Monte Carlo code has a very large statistical error (see error bars)
because the polarization modulus is small.}. We see that the
polarization decreases rapidly in the IR and remains parallel to the disk.
Therefore the rotation is not observed in model 4. The same result is seen for
amorphous carbon models of Carciofi, Magalh\~aes, \& Bjorkman (2002). The
polarization modulus in NIR provided by the
code must be seen with caution because the model does not include the dust
(re)emission, which may be important at those wavelengths.

The absence of angle rotation in model 4 is caused by the large
absorption of the amorphous carbon. If a less absorptive
material was used (as silicate), the rotation could be produced. So the
polarization angle restricts the envelope optical depth and also
the index of refraction of the dust. This means that JHK polarization
measurements would be useful to better determine the grain composition in the
envelope of \hhe.

A more complete study of the dust torus of \he using a code that includes
the radiative transfer and the thermal equilibrium is being carried on
(C.~V. Rodrigues \& A.~C. Carciofi, in preparation). The flux from optical to IR
wavelengths, as well as the polarization, will be modeled. This will put
more constraints on the physical properties of the dust envelope of \hhe.

\section{Conclusions}

In this work we analyze BVRI polarization and IR photometry of the
proto-planetary nebula Hen 3-1475. The main results are summarized below:

\begin{itemize}

\item the optical broad band fluxes of \he are highly polarized.
A large fraction of this polarization may be produced in the circumstellar
environment of this object;

\item the BVRI polarization position angle is perpendicular to the axis of the
bipolar structure seen in optical images. As the aperture
used to measure the polarization is small and therefore does not include the
jets, that implies that the central disk must be optically thick in all the
bands observed or the dust must be composed of a highly absorptive material;

\item the wavelength dependence of the polarization of \he was modeled using a
Monte Carlo radiative transfer code. We could find a good fit using
an index -3.0 for the power-law grain-size distribution, whose limits are 0.063 and 0.22
$\mu$m. In spite of the claim for an oxygen chemistry envelope (Knapp et al.
2000), the optical constants of amorphous carbon produce a better fit to 
the optical polarization than those of silicates. However, the simple model
used here does not allow to discard silicate dust in the \he circumstellar envelope;

\item our CCD images provide us with polarimetric data of a large number of
stars in the \he field. So we could reliably estimate the foreground
polarization in this direction. Curiously, the foreground polarization has the
same direction of the intrinsic polarization of \hhe. This might be
interpreted as an interplay between the IS magnetic field and the \he
geometry;

\item the IR colors of \he suggests an interstellar reddening in the V
band of around 3.2 mag in the line of sight. The circumstellar extinction
was estimated as A(V) $\approx$ 9.0 mag.

\end{itemize}



\acknowledgments

We are grateful to A. Damineli for lending us observational time and to
Alex C. Carciofi for suggestions to the manuscript. We also acknowledge the 
referee for the suggested improvements.
JGH thanks FAPESP Proc. No. 2001/09018-2. 

This publication makes use of data products from the Two Micron All Sky
Survey, which is a joint project of the University of Massachusetts and the
Infrared Processing and Analysis Center/California Institute of Technology,
funded by the National Aeronautics and Space Administration and the National
Science Foundation.

This research has also made use of NASA's Astrophysics Data System Bibliographic
Services and of the SIMBAD database, operated at CDS, Strasbourg, France.




\clearpage


\clearpage

\begin{deluxetable}{ccccccc}
\tablecaption{Polarization of Hen~3-1475 and of field stars
\label{tab-pol}}
\tablewidth{0pt}
\tablehead{
\colhead{Object} &
\colhead{Band} &
\colhead{$P$} &
\colhead{$\sigma$} &
\colhead{$\theta$} &
\colhead{$\sigma$} &
\colhead{Number of} \\
  &  &
\colhead{(\%)} &
\colhead{(\%)} &
\colhead{($^o$)} &
\colhead{($^o$)} &
\colhead{observations}}
\startdata
Hen~3-1475 & B & 9.12 & 0.65 & 43.7 & 2.0 & 1 \\
(Observed) & V & 8.31 & 0.28 & 43.1 & 1.0 & 2\tablenotemark{a}\\
           & R$_C$ & 7.17 & 0.29 & 41.3 & 1.2 & 1 \\
           & I$_C$ & 5.70 & 0.20 & 40.2 & 1.0 & 1 \\
\\
Hen~3-1475  & B & 7.81 & 0.65 & 43.7 & 2.4 & - \\
(Intrinsic) & V & 7.03 & 0.28 & 42.1 & 1.1 & - \\
            & R$_C$ & 5.95 & 0.29 & 40.2 & 1.4  & - \\
            & I$_C$ & 4.71 & 0.20 & 38.4 & 1.2 & - \\
\\
Field  & B & 1.313 & 0.032 & 43.45 & 0.7 & 95 \\
Stars  & V & 1.298 & 0.009 & 48.30 & 0.2 & 501 \\
       & R$_C$ & 1.243 & 0.008 & 46.25 & 0.2 & 617 \\
       & I$_C$ & 1.041 & 0.006 & 47.99 & 0.2 & 901 \\
\hline
\enddata
\tablenotetext{a}{average of the two observations}
\end{deluxetable}

\clearpage

\begin{deluxetable}{cccccccc}
\tablecaption{Serkowski law fits to the observed polarization of Hen~3-1475 and
to the field stars (K = 1.66 $\lambda_{max}$ + 0.01 - Whittet et al. 1992)
\label{tab-serk}}
\tablewidth{0pt}
\tablehead{
\colhead{Object} &
\colhead{$P_{max}$} &
\colhead{$\sigma$} &
\colhead{$\lambda_{max}$} &
\colhead{$\sigma$} &
\colhead{$K$} &
\colhead{$\sigma$} &
\colhead{$\chi^2$} \\
 &
\colhead{(\%)} &
\colhead{(\%)} &
\colhead{($\mu m$)} &
\colhead{($\mu m$)}
}
\startdata
Hen~3-1475 & 10.8 & 3.3 & 0.24 & 0.18 & 0.40 & 0.30 & 0.74 \\
Field stars & 1.328 & 0.012 & 0.4746 & 0.0092 & 0.798 & 0.015 & 5.4 \\
\enddata
\end{deluxetable}

\clearpage

\begin{deluxetable}{lcc}
\tablecaption{Input parameters for the polarization model of \he
\label{tab-input}}
\tablewidth{0pt}
\tablehead{
\colhead{Input parameter} &
\colhead{Fixed} &
\colhead{Value}
}
\startdata
Source radius & Yes & 20 R$_\odot$\tablenotemark{a}\\
Internal radius of the disk, $R_i$ & Yes & $4\ \times\ 10^5$
R$_\odot$\tablenotemark{b}\\
External radius of the disk & Yes & $1.7\ \times\ 10^6$
R$_\odot$\tablenotemark{c} \\
Height of the disk, $h$ & No & 4 - 9 $\times 10^5$ R$_\odot$ \\
Inclination angle between disk axis and plane of sky & Yes &
40$\arcdeg$\ \tablenotemark{d}\\
Optical depth in V band, $\tau_V$ & No & Fitted \\
Mininum limit of the size distribution, $a_{min}$ & No & Fitted \\
Maximum limit of the size distribution, $a_{max}$&  No & Fitted \\
Exponent of the size distribution, $q$ & No & -3.5 to -2.5 \\
Refractive index & Yes & Literature\tablenotemark{e} \\
\enddata
\tablenotetext{a}{Estimated in this work}
\tablenotetext{b}{From the IRAS fluxes (Bobrowski et al. 1995)}
\tablenotetext{c}{From optical image (Riera et al. 1995; Bobrowski et al. 1995)}
\tablenotetext{d}{From optical spectroscopy (Borkowski \& Harrington 2001)}
\tablenotetext{e}{See text}

\end{deluxetable}

\clearpage

\begin{deluxetable}{cccccccc}
\tablecaption{Parameters of the models for the spectral dependence of the
polarization of \he
\label{tab-model}}
\tablewidth{0pt}
\tablehead{
\colhead{Model} &
\colhead{Material} &
\colhead{$q$} &
\colhead{$h/2$} &
\colhead{$\tau_{V}$} &
\colhead{$a_{min}$} &
\colhead{$a_{max}$} &
\colhead{$\chi^2$} \\
 &  &  &
\colhead{(R$_\odot$)} & &
\colhead{($\mu m$)} &
\colhead{($\mu m$)}
}
\startdata
1 & Amorphous carbon - Z96 & -3.5 & $8 \times 10^5$ & 9.0 & 0.063 & 0.21 & 13.0
\\
2 & Amorphous carbon - Z96 & -3.0 & $8 \times 10^5$ & 9.0 & 0.064 & 0.21 & 7.5
\\
3 & Amorphous carbon - Z96 & -2.5 & $8 \times 10^5$ & 9.0 & 0.062 & 0.21 & 11.6
\\ \\
4 & Amorphous carbon - Z96 & -3.0 & $7 \times 10^5$ & 12.5 & 0.063 & 0.22 & 2.5
\\
\\
5 & Silicate O-rich - O92 & -3.5 & $8 \times 10^5$ & 9.0 & 0.072 & 0.26 & 30.7
\\
6 & Silicate O-rich - O92 & -3.0 & $8 \times 10^5$ & 9.5 & 0.070 & 0.26 & 19.2
\\
\\
7 & Astronomical silicate - DL84 &-3.5 & $8 \times 10^5$ & 9.0 & 0.064 & 0.29 &
39.9 \\
\enddata
\end{deluxetable}

\end{document}